\documentclass[twocolumn]{article}

\newcommand{\beq}{\begin{equation}}
\newcommand{\eeq}{\end{equation}}
\newcommand{\beqa}{\begin{eqnarray}}
\newcommand{\eeqa}{\end{eqnarray}}
\newcommand{\Eq}[1]{Eq.\ (\ref{#1})}
\newcommand{\Eqs}[2]{Eqs.\ (\ref{#1}) and (\ref{#2})}
\newcommand{\Fig}[1]{Fig.\ \ref{#1}}
\newcommand{\Figs}[2]{Fig.\ \ref{#1} and \ref{#2}}

\title{Linear model of tumor growth in a changing environment}
\author{Jos\'e F. Nieves\footnote{nieves@ltp.upr.clu.edu} and
Marcelo R. Ubriaco\footnote{ubriaco@ltp.upr.clu.edu}\\
Laboratory of Theoretical Physics\\
Department of Physics \\
P.O. Box 23343, Rio Piedras\\
Puerto Rico 00931-3343, USA}

\begin{document}
\maketitle

\begin{abstract}
We propose a model for describing the growth on an untreated tumor,
which is characterized in a simple way by a minimal number of parameters
with a well-defined physical interpretation.
The model is motivated by invoking the \emph{Master Equation} and the
\emph{Principle of Detailed Balance} in the present context,
and it is easily generalizable to include the effects of various
types of therapies. In the simplest version that we consider here,
it leads to a linear equation that describes
the population growth in a dynamic environment,
for which a complete solution can be given in terms of the integral
of the growth rate. The essential features of the general solution
for this case are illustrated with a few examples.

\emph{Keywords}: Cancer, Mathematical model
\end{abstract}

\section{Introduction}
\label{sec:intro}

In the last few decades, the mathematical modeling of tumor growth
as a function of time has been based mainly in applications
of the Gompertz equation\cite{laird,norton1,norton2} and a modified
version of it\cite{calderon}, power law equations\cite{calderon,hart},
and generalizations of the logistic equation\cite{spratt,dingli}.
These mathematical models helped to provide an understanding of
tumor growth 
as a more organized phenomenon than previously thought. 
In particular, solutions of the Gompertz 
and the modified logistic equations exhibit an S-shaped form which represents
inhibition of growth to an asymptotic limit. 
In addition, their
applications to experimental data, with the help of additional
differential equations to model therapy, has served as a guide 
to improve the effectiveness of treatment.

However, a preference for one model or the other has been
based exclusively on their adequacy to fit experimental data.
Similarly, some of the models consist on generalizing previous ones
by modifying a parameter in a way that is not motivated
by any fundamental principle, but again to fit some data.
For example, the so-called \emph{Generalized Logistic} model
is the result of merely
modifying the logistic equation by inserting an arbitrary power which,
in order to fit the data, becomes a noninteger number. 

While this may be appropriate for particular purposes, and for applications
to systems that are not governed by the laws of Nature, the fact
that the models are neither motivated nor based on 
fundamental physical principles prevents us from ascribing
a physical meaning to the parameters that appear in such models.
This makes it more difficult, if not impossible, to establish
a connection between the observable effects that can be described
on the basis of phenomenological models,
with more fundamental explanations and understandings of
the mechanism of growth in biological systems, which must be based
on detailed microscopic dynamics.

The model is
motivated by invoking the \emph{Master Equation} and the Principle
of Detailed Balance in the present context. In the simplest version
that we consider here, it leads to a linear equation that describes
the population growth in a dynamic environment,

The main objective of this article is to present a
model that is inspired by fundamental physical principles, such as
the \emph{Master Equation} and the Principle
of Detailed Balance, and is partly motivated by analogous
equations that apply to a variety of physical systems. The model is based on
a linear differential equation with a time dependent coupling.
It shares the same asymptotic behavior of the GL model,
giving an S-shaped function for tumor growth that vanishes at $t = 0$
and eventually reaches an asymptotic value at large $t$. 
However, the model proposed here contains a minimal number of parameters,
which have a concrete and well-defined meaning and
are in principle determined and calculable if the interactions
that govern the underlying microscopic mechanism of growth are known.
By the same token,
the use of this model should in turn shed light on such mechanisms,
thereby providing a firm footing for pursuing and extending such approaches.

The rest of this paper is organized as follows. In Section\ \ref{sec:previous},
for the purpose of comparing our proposed
differential equation with the previously proposed models,
we briefly recall the basic aspects and shortcomings of them.	
In Section\ \ref{sec:themodel} we discuss the theoretical framework
that motivates the model that we propose, the
assumptions and idealizations involved, and the interpretation
of the parameters that appear in it. In Section\ \ref{sec:solution}
the general solution to the equation is displayed, and it is illustrated
by considering various specific examples that can be of practical
use. Finally, Section\ \ref{sec:conclusions} contains our conclusions.

\section{Previous models}
\label{sec:previous}

We will denote by $f(t)$ the growth function, and for simplicity of the
notation omit the argument $t$ except when necessary to avoid
confusion. We envisage $f$ to be the number of cells at a 
particular instant of time.

The Gompertz equation can be written in the form
\beq
\label{Gompertz}
\frac{df}{dt} = -\alpha f \ln\left(\frac{f}{\beta}\right)\,,
\eeq
where $\alpha$ and $\beta$ are two parameters. The solution is given by
\beq
f(t) = \beta e^G\,,
\eeq
with
\beq
\label{GompertzG}
G \equiv \left[\ln\left(\frac{f_0}{\beta}\right)\right]e^{-\alpha t} \,,
\eeq
where $f_0$ is the initial value of $f$. The
asymptotic value is determined as
\beq
f_\infty = \beta \,.
\eeq
A particular feature of \Eq{Gompertz} is that it is not defined for $f = 0$,
and therefore it does not allow the initial value $f_0 = 0$, as
\Eq{GompertzG} reveals, thus ruling out its
application to data with rather small initial population value.

A way to overcome this difficulty
is to modify \Eq{Gompertz} by introducing an arbitrary
parameter $\epsilon > 0$ such that the new differential equation
reads\cite{calderon}
\beq
\frac{df}{dt}=\alpha f \left[\ln\left(\frac{\beta}{f}\right)\right]
^{1+\epsilon}\,,
\eeq
which has a non trivial solution given by
\beq
f(t) = \beta e^{-G^\prime} \,,
\eeq
where
\beq
G^\prime \equiv \left(\frac{1}{\epsilon\alpha t}\right)^{\frac{1}{\epsilon}}
\,.
\eeq
While this procedure overcomes the above-mentioned difficulty with the Gompertz
solution, a physical motivation or justification is lacking.

Another approach\cite{hart} has been to consider the power law differential
equation
\beq
\frac{df}{dt}=\beta f^{\alpha}\,.
\eeq
For $\beta\neq 1$ it has the solution
\beq
f(t) = \left[\beta t (1-\alpha) + f_0^{1 - \alpha}\right]^
{\frac{1}{1-\alpha}}\,,
\eeq
which leads to linear or exponential growth for
$\alpha = 0$ or $\alpha = 1$, respectively. Certainly, power law growth is
unconstrained, and its behavior is radically different from that
given by \Eq{Gompertz}.

Another way to obtain asymptotic behavior in tumor growth is to modify
the logistic equation by introducing an arbitrary power $\epsilon > 0$,
leading to the differential equation
\beq
\label{GL}
\frac{df}{dt} = \frac{\beta}{\epsilon} f\left[1 - 
\left(\frac{f}{\alpha}\right)^{\epsilon}\right]\,,
\eeq
whose solution is given by
\beq
\label{glsol}
f(t) = \frac{f_0}{L^{1/\epsilon}} \,,
\eeq
where
\beq
L \equiv 
\left(\frac{f_0}{\alpha}\right)^\epsilon + 
\left[
1 - \left(\frac{f_0}{\alpha}\right)^\epsilon\right]e^{-\beta t}\,.
\eeq
\Eq{glsol} reproduces the Gompertz function for $\epsilon = 0$, but it also
has the limitation that 
the differential equation \Eq{GL} does not allow a zero value at $t=0$.
Therefore,
the solution of \Eq{GL} can be considered to be the most general function that
possesses
asymptotic behavior and a shape consistent with experimental data.
However, the fact that \Eq{GL}
can be neither motivated nor understood on the basis
of some of fundamental physical principles is in our opinion
an important limitation to further understanding the mechanisms
that drive the growth in these biological systems, on the basis
of such equations.
The main reason for this lack
of insight is due to the fact that \Eq{GL} is the result of merely
modifying the logistic equation by inserting an arbitrary power, that in order
to fit the data becomes a noninteger number.

Therefore, without any other justification
it is not possible in this and the previously discussed models
to ascribe a deeper physical significance to the parameters of the model.
	
\section{The Model}
\label{sec:themodel}

Our starting point is the equation that describes the growth of a population
that is sustained by an environment. We assume that in such situations
the population grows up to a certain saturation limit $f_{s}$, and that the
environment is large enough such that it is not affected by the population
itself. Under such conditions, we assert that the rate of change
of the population is proportional to the difference between actual
value of the population and its saturation limit. Therefore,
\beq
\label{rateeq}
\dot{f} = -\gamma(f - f_{s})\,.
\eeq
This equation is reminiscent of Newton's cooling law which states
that the rate of change of the temperature of a system is proportional
to the deviation of the system's temperature from the temperature of
its environment.  In our context it is possible and useful to give a motivation
and justification in terms of more basic principles as follows.

\subsection{The Master Equation}

The problem of the time evolution of the population of a given
specie appears in many physical contexts. For example, in
the astrophysical context of the \emph{Early Universe}, one analogous
problem is the determination of the abundance of the various atomic elements
and how they form\cite{bernstein}.
The nucleosynthesis processes in stars are examples
of similar phenomena which, among other things, explain the generation
of energy in the Sun.

In one way or another, the two basic principles that guide the development
of a population are the Master Equation and the Principle of Detailed
Balance\cite{huang,tolman}. The master equation takes the form
\beq
\frac{df}{dt} = W \,,
\eeq
where $W$ depends on $f$ itself and the other variables that describe the
rest of the system with which the population can interact. $W$ is decomposed
into a series of terms, each of which represents the contribution
due to a particular process that causes the population to change.
The principle of detailed balance states that there is a precise
relation between the so-called direct process and its inverse.

For example, let us consider a process in which only one cell participates
and let us denote such process in symbols by
\beq
\phi \leftrightarrow X\,,
\eeq
where $\phi$ stands for a member of the population (a cell) and $X$
stands for a different object. In the direct process, indicated by
the right-pointing arrow, a cell $\phi$ disappears into $X$, while in the 
inverse process, indicated by the left-pointing arrow,
the reverse is true. Then, denoting by
$\gamma_d$ and $\gamma_i$ the rates for the direct and inverse processes,
respectively, their contribution to $W$ is written in the form
\beq
\label{W1}
W_1 = -\gamma(f - f_s) \,,
\eeq
where
\beq
\gamma = \gamma_d - \gamma_i \,.
\eeq

Similar equations also describe the kinetic approach
to equilibrium of systems
that are put in contact with a reservoir. In such cases, which are
governed by physical kinetic equations, the principle of
detailed balance implies a fundamental relation
\beq
\gamma_i = e^{-\Delta E/T} \gamma_d \,,
\eeq
where $\Delta E$ and $T$ are identified with the change in energy of the
system and the temperature of the environment, respectively.

In our case, in principle both $\gamma_{d,i}$ could be calculated
if the interaction between the cells with their surroundings and among
themselves were known.
However, we have at present no such theory of these interactions.
Thus, we leave $\gamma$ as an unknown parameter with the
property that is a positive quantity.

The procedure outlined above for the case of single cell processes can
be generalized to more complicated ones. For example, consider the processes
in which two cells participate, which we denote in symbols by
\beq
\phi\phi \leftrightarrow X \,.
\eeq
Because the direct process involves two cells, its rate is proportional
to $f^2$. By the same reasoning that lead us to write \Eq{W1},
the contribution from these processes to $W$ is of the form
\beq
W_2 = -\gamma^\prime (f^2 - f^2_s) \,,
\eeq
where $\gamma^\prime$ characterizes the rate for the process to occur.

As a typical rule in those contexts in which these equations have already been
applied, the processes in which more than two members participate
are rare and not important. Therefore, we are tempted to state that
the master equation
\beq
\label{me}
\frac{df}{dt} = -\gamma(f - f_s) - \gamma^\prime (f^2 - f^2_s) \,,
\eeq
is a good starting point for further exploration of these ideas
in the present context as well.

In the present paper, we will restrict ourselves to the linear term
only, as written in \Eq{rateeq}. The assumption behind this approximation is
that the process in which the cells participate in pairs are rare
compared to those in which only one cell participates. Should this linear
approximation prove to be inadequate, it could indicate that the
pair interactions are important and the quadratic terms in \Eq{me}
should be taken into account. Overall, this approach provides a
framework for carrying a systematic analysis, based on incremental
approximations, on a firm footing and in an organized fashion.

\section{Solution}
\label{sec:solution}

\subsection{Static environment}

When $\gamma$ is a constant, \Eq{rateeq} has the simple solution
\beq
\label{fsolstatic}
f(t) = f_s\left[1 - e^{-\gamma t}\right] + f_0 e^{-\gamma t} \,,
\eeq
where $f_0$ is the initial population, which can of course be taken to be zero.
However, notice that the population reaches the saturation limit $f_{s}$ 
independently of the initial value $f_0$.  This contrasts with the solution
of the GL model\cite{dingli},
which requires a non-zero value $f_0$ or otherwise the
solution is the trivial solution $f(t) = 0$.

\subsection{Dynamic environment}

We consider the case in which the environment can change
due to external influences. For us this means that the parameters
$\gamma$ and $f_s$ that appear in the model equation, both of
which depend on the state of the environment, change with time.
In the absence of a dynamical theory of the interactions of the cells,
all we can do is promote $\gamma$ and $f_s$ to be functions of time.
Therefore, our basic equation becomes
\beq
\label{rateeqt}
\frac{df}{dt} = -\gamma(t)\left[f - f_{s}(t)\right]\,.
\eeq
This equation is conveniently solved by the Green function method,
\beq
f(t) = \int^\infty_0 dt^\prime G(t,t^\prime)\gamma(t^\prime)f_s(t^\prime) +
f_h(t)\,,
\eeq
where $G$ satisfies
\beq
\label{Geq}
\frac{dG}{dt} + \gamma G = \delta(t - t^\prime) \,,
\eeq
$f_h$ is a solution to the homogeneous equation such that $f$ satisfies
the initial condition $f(0) = f_0$. In \Eq{Geq}, $\delta(x)$ stands
for the Dirac delta function.
A suitable Green function for \Eq{rateeqt} is
\beq
G(t,t^\prime) = \theta(t - t^\prime)g(t,t^\prime) \,,
\eeq
where $g(t,t^\prime)$ is the solution to the homogeneous equation satisfying
the condition $g(t^\prime,t^\prime) = 1$, and $\theta(x)$ is the unit
step function. The function $g(t,t^\prime)$ is then
uniquely determined as
\beq
\label{g}
g(t,t^\prime) \equiv e^{-\int^t_{t^{\prime}}\,dt^{\prime\prime}
\gamma(t^{\prime\prime})} \,,
\eeq
and therefore the solution for $f(t)$ is given by
\beq
\label{fsol}
f(t) = \int^t_0 dt^\prime\, g(t,t^\prime)\gamma(t^\prime)f_s(t^\prime) +
f_0\,g(t,0) \,.
\eeq
Needless to say, if $\gamma$ and $f_s$ are assumed to be constant,
then \Eq{fsol} reduces to the solution given in \Eq{fsolstatic}. But
for any imaginable functions $\gamma(t)$ and $f_s(t)$ that could
be used to parametrize the changing environment,
\Eq{fsol} readily provides the complete solution in terms of two integrals.

A particularly simple form of the solution is obtained in the case
in which $f_s$ is a constant. In this case, noting from \Eq{g} that
\beq
\frac{dg(t,t^\prime)}{dt^\prime} = \gamma(t^\prime)g(t,t^\prime) \,,
\eeq
then \Eq{fsol} yields
\beq
\label{fsolfsconst}
f(t) = f_s\left[1 - g(t,0)\right] + f_0\,g(t,0)\,,
\eeq
where we have used $g(t,t) = 1$.

\subsection{Examples} 

In order to illustrate some general features of the solution, we will
consider below various specific cases.

\subsubsection{Example 1}

Let us assume that $f_s$ is a constant, while
$\gamma$ varies as some (integer) power of $t$; i.e.,
\beqa
\label{gammaex1}
f_s & = & \mbox{constant} \nonumber\\
\gamma & = & a t^n \,,
\eeqa
where $a$ is positive constant and $n$ is a positive integer.
First, from \Eq{g},
\beq
g(t,t^\prime) = e^{-\gamma_0\left(t^{n + 1} - t^{\prime\, n + 1}\right)} \,,
\eeq
where we have defined 
\beq
\label{gamma0}
\gamma_0 = \frac{a}{n + 1} \,,
\eeq
for simplicity of the notation. The solution obtained from \Eq{fsolfsconst} is
then  
\beq
\label{fsolpower}
f(t) = f_s\left[1 - e^{-\gamma_0 t^{n + 1}}\right]
+ f_0e^{-\gamma_0 t^{n + 1}} \,.
\eeq

\begin{figure}
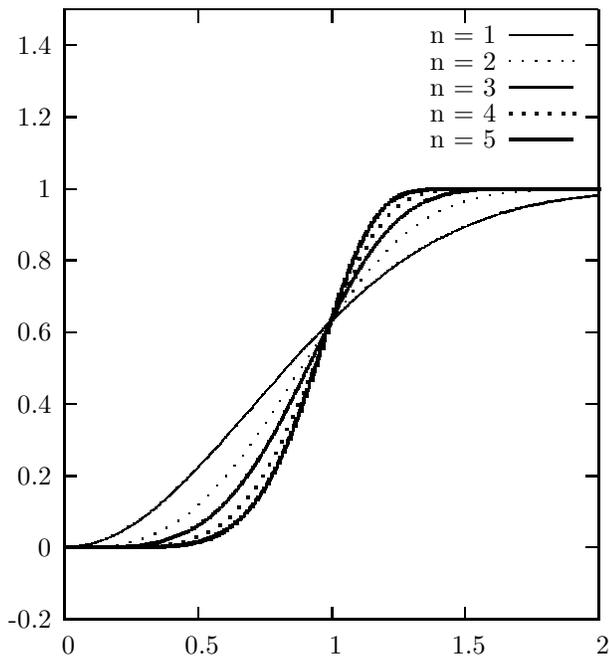

\begin{center}
\setlength{\unitlength}{0.240900pt}
\ifx\plotpoint\undefined\newsavebox{\plotpoint}\fi
\sbox{\plotpoint}{\rule[-0.200pt]{0.400pt}{0.400pt}}%

\end{center}
\caption[]{Plot of $f(t)/f_s$ with $f(t)$ given by \Eq{fsolpower},
with the initial condition $f_0 = 0$,
for $\gamma = a t^n$ with $n = 1,2,3,4,5$, in arbitrary time units.
\label{fig:ex1}
}
\end{figure}

In \Fig{fig:ex1} we plot the function $f/f_s$,
in arbitrary time units (i.e., setting $\gamma_0 = 1$), for
the values of the exponent $n = 1,2,3,4,5$, and taking $f_0 = 0$.

\subsubsection{Example 2}

\begin{figure}
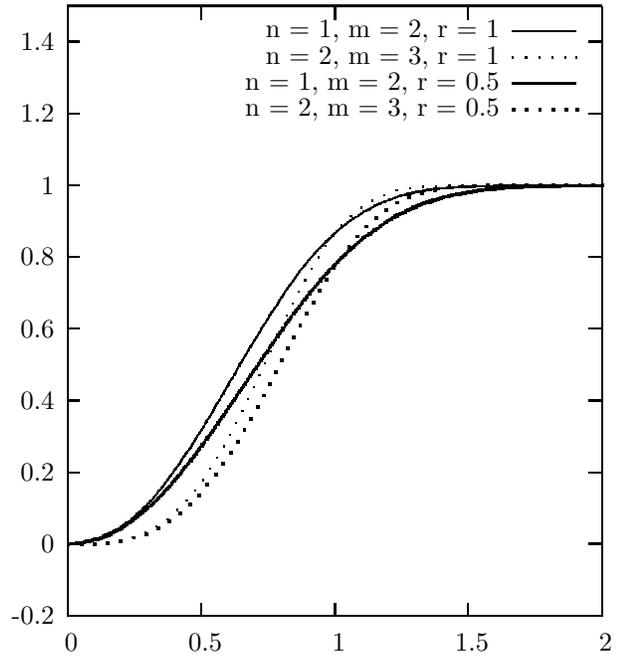

\begin{center}
\setlength{\unitlength}{0.240900pt}
\ifx\plotpoint\undefined\newsavebox{\plotpoint}\fi
\sbox{\plotpoint}{\rule[-0.200pt]{0.400pt}{0.400pt}}%

\end{center}
\caption[]{Plot of $f(t)/f_s$ with $f(t)$ given by \Eq{fsolpower2},
with the initial condition $f_0 = 0$, in arbitrary time units,
for various values of $n$, $m$ and the ratio $r = \gamma_0/\gamma^\prime_0$.
\label{fig:ex2}
}
\end{figure}

As before, we assume that $f_s$ is a constant, but now
take $\gamma$ as a combination of two monomials; i.e., 
\beqa
\label{gammaex2}
f_s & = & \mbox{constant} \nonumber\\
\gamma & = & a t^n + b t^m \,.
\eeqa
where $n,m$ are positive integers and $a,b$ are positive constants.
Following the same steps as above, the obtained from \Eq{fsolfsconst}
is given by
\beq
\label{fsolpower2}
f(t) = f_s\left[1 - e^{-(\gamma_0 t^{n + 1} + \gamma^\prime_0 t^{m + 1})}
\right] + f_0e^{-(\gamma_0 t^{n + 1} + \gamma^\prime_0 t^{m + 1})} \,,
\eeq
where $\gamma_0$ is defined as in \Eq{gamma0} and in analogous fashion
\beq
\gamma^\prime_0 \equiv \frac{b}{m + 1} \,.
\eeq
In \Fig{fig:ex2} we plot the function $f/f_s$, for various values
of $n$, $m$ and the ratio $r = \gamma_0/\gamma^\prime_0$.

We stress that, apart from the initial and saturation values $f_0$ and
$f_s$, respectively, the only unknown and ajustable parameters
of these models are the constants that appear in the ansatz for the
growth rate function $\gamma(t)$, e.g. the constants $a$ and $b$ defined in
\Eqs{gammaex1}{gammaex2}.

\subsection{Generalizations}

By inspection,
\Eq{fsolpower2} can be generalized in an obvious way to the case
in which $\gamma(t)$ is a power series in $t$. Although the models
considered above, with $f_s$ taken to be a constant, already provide
a flexible and rich structure as far as their ability to fit
the phenomenlogical data is concerned, it is possible to consider the cases
in which $f_s$ is a function of $t$. In these cases it is not
possible to give a closed expression for the solution given in
\Eq{fsol}, in general. However, in other analogous physical problems where
similar situations arise\cite{bernstein}, very effective approximation methods
have been used which could be employed in these cases as well.

\section{Conclusions and Outlook}
\label{sec:conclusions}

In this article we have presented a model for tumor growth, which is
based on physical principles on one hand, together with
plausible physical assumptions and idealizations on the other.
As shown in Section\ \ref{sec:solution}, in the simplest
version of the model, in which the pairwise interactions between
the cells are neglected, the growth equation is linear and a
complete solution can be readily given. Moreover, the 
solutions were explicitly given for a few sample cases, which
exhibit the known characteristic features of tumor growth.
Thus the approach that we have followed is fruitful in several ways.
Firstly, the model contains a minimal number of parameters, which
have a concrete and well-defined meaning, and
are in principle determined and calculable if the interactions
that govern the underlying microscopic mechanism of growth are known.
Secondly, by the same token,
the use of this model should in turn shed light on such mechanisms,
thereby providing a firm basis for pursuing this line of work.
Thirdly, the model can be extended beyond the linear approximation that we have
used if the
pairwise interactions are believed to be important in a particular system
and the quadratic terms in \Eq{me} should be taken into account.
Our approach provides a
framework for taking into account such higher order terms
in a systematic fashion. Lastly, while in this paper we have restricted
ourselves to treat the growth of an untreated tumor, our work
paves the way for applying similar principles and ideas
to include the effects of therapy. Work along these lines is in progress.

\section*{Acknowledgments}
This material is based upon work supported by the US National
Science Foundation under Grant No. 0139538 (JFN).

\end{document}